\documentclass[aps,prl,twocolumn,showpacs,nofootinbib,preprintnumbers,superscriptaddress,letterpaper,amsmath,amssymb]{revtex4}

\usepackage{graphicx}  
\usepackage{dcolumn}   
\usepackage{bm}        
\usepackage{amssymb}   
\usepackage{feynmf}
\usepackage{slashed}
\def\gsim{\lower0.5ex\hbox{$\:\buildrel >\over\sim\:$}}
\def\lsim{\lower0.5ex\hbox{$\:\buildrel <\over\sim\:$}}

\newcommand{\be}{\begin{equation}}
\newcommand{\ee}{\end{equation}}
\newcommand{\bea}{\begin{eqnarray}}
\newcommand{\eea}{\end{eqnarray}}

\newcommand{\nbox}{{\,\lower0.9pt\vbox{\hrule \hbox{\vrule height 0.2 cm
\hskip 0.2 cm \vrule height 0.2 cm}\hrule}\,}}

\begin{document}

\thispagestyle{empty}
\vspace*{-3.5cm}

\vspace{0.5in}

\title{Collider searches for dark matter in events with a $Z$ boson and missing energy}

\begin{center}
\begin{abstract}
Searches for dark matter at colliders typically involve signatures with
 energetic initial-state radiation without visible recoil
particles.  Searches for mono-jet or mono-photon signatures have yielded powerful
constraints on dark matter interactions with Standard Model
particles. We extend this to the mono-$Z$ signature and reinterpret an ATLAS
analysis of events with a $Z$ boson and missing transverse momentum to
derive constraints on dark matter interaction mass scale and nucleon
cross sections in the context of effective field theories describing dark matter which
interacts via heavy mediator particles with quarks or weak bosons.
\end{abstract}
\end{center}

\author{Linda M. Carpenter}
\affiliation{Department of Physics The Ohio State University, Columbus, OH 43210}
\affiliation{Department of Physics and Astronomy, University of California, Irvine, CA 92697}
\author{Andrew Nelson}
\author{Chase Shimmin}
\author{Tim M.P. Tait}
\author{Daniel Whiteson}
\affiliation{Department of Physics and Astronomy, University of California, Irvine, CA 92697}
\preprint{UCI-HEP-TR-2012-21}
\pacs{95.35.+d, 14.70.Bh}
\maketitle

The particle nature of dark matter is one of the greatest outstanding
mysteries of cosmology and particle physics.  A suite of dedicated
experiments seek to shed light on how dark matter interacts with Standard
Model (SM) particles by looking for detection of ambient dark matter
either directly scattering with heavy nuclei or indirectly through its annihilation
into high energy Standard Model particles.
An important third pillar to the search for the particle nature of dark matter is furnished
by high energy particle accelerators, which can produce pairs of dark matter particles,
which are expected to manifest as an excess of events showing an
imbalance in momentum conservation.
Searches for missing transverse momentum are a major activity at the LHC precisely because
of their potential connection to dark matter \cite{Morrissey:2009tf}.

Searches for dark matter in missing momentum channels can be classified based on the visible particles
against which the invisible particles recoil.  Existing experimental studies have considered cases in
which the visible radiation is a jet of hadrons (initiated by a quark or
gluon)~\cite{cdfjmet,ATLAS:2012ky,Chatrchyan:2012me}, a photon~\cite{Aad:2012fw,Chatrchyan:2012tea}, or
a $W$ boson decaying into leptons~\cite{Bai:2012xg}.  These studies are
performed in the context of
an effective field theory (EFT) which captures the physics of a heavy particle mediating
an interaction between dark matter and quarks and/or gluons.
In this article, we extend the menu of such searches to include the case of a $Z$ boson decaying
into a pair of charged leptons (electrons or muons) and recast the recent ATLAS measurement
of $ZZ\rightarrow \ell\ell\nu\nu$~\cite{atlaszz} into a bound on production of dark matter
in association with a $Z$ boson\footnote{This signature has previously been considered in the context of a collider search for $Z^\prime$ decaying to invisible modes (Ref.~\cite{ZprimeDM}), and more recently in the dark matter
context with a slightly more model-dependent framework in Ref.~\cite{Bell:2012rg}.}.  Since our signature consists
of a pair of leptons consistent with a $Z$ boson decay recoiling against transverse momentum carried
by particles invisible to the detector, we refer to our selection as a ``mono-$Z$" signature.

We work in the context of
EFTs where the dark matter's primary interactions are with quarks or directly
with electroweak bosons.  In the case where interactions are primarily with quarks,
the mono-$Z$ signature arises from a $Z$ boson which is radiated from a
$q \bar{q}$ initial state, much like mono-jets or mono-photons.  Such interactions
also imply (depending on the specific form of the interaction)
large rates for scattering with heavy nuclei.
The case of direct interactions with a pair of $Z$ bosons is more challenging to connect to
direct detection (however, see \cite{Frandsen:2012db}),
and is a particular strength of collider searches.  As usual, in both
cases the EFT will break down at an energy not far above the one which characterizes
the strength of the interactions written as higher dimensional operators, and is only
a good description of the physics for processes taking place at energies well below
this cut-off scale.

\section{Effective Field Theory}

\subsection{Interactions with Quarks}

Effective field theories for dark matter interacting primarily with SM quarks
have been considered in Refs.~\cite{Beltran:2008xg,Shepherd:2009sa,Cao:2009uw,Beltran:2010ww,Goodman:2010yf,Bai:2010hh,Goodman:2010ku,Rajaraman:2011wf,Fox:2011pm,Cheung:2012gi}.  We consider the interactions,
\bea
 &  \sum_q & \left\{
\frac{m_q}{\Lambda_{\rm D1}^3} \bar{q} q~ \bar{\chi} \chi
+ \frac{1}{\Lambda_{\rm D8}^2}  \bar{q} \gamma^\mu \gamma_5 q~
\bar{\chi} \gamma_\mu \gamma_5 \chi \right.
\nonumber \\ & & \left.
+ \frac{1}{\Lambda_{\rm D5}^2}  \bar{q} \gamma^\mu q~ \bar{\chi} \gamma_\mu \chi
+ \frac{1}{\Lambda_{\rm D9}^2}  \bar{q} \sigma^{\mu \nu} q~
\bar{\chi} \sigma_{\mu \nu} \chi
\right\}
\label{eq:EFTq}
\eea
where $\chi$ is the dark matter particle, which we assume to be a Dirac fermion,
$q$ is a SM quark, and the coefficients $\Lambda$ parameterize the coupling
strength of scalar (D1), vector (D5), axial-vector (D8), and tensor (D9)
interactions between the two.  The labeling scheme is adopted from
Ref.~\cite{Goodman:2010ku}, with the choices dictated as those operators
which lead to a non-vanishing scattering rate with nucleons at small momentum transfer.
We will typically consider one interaction type to
dominate at a time, and will thus keep one $\Lambda$ finite while the rest
are sent to infinity and decoupled.  These operators are normalized so as to be
consistent with minimal flavor violation.

\subsection{Interactions with $Z$ Bosons}

One may also construct an EFT in which the dark matter interacts directly with pairs of electroweak bosons.  Given
our assumption that $\chi$ is a SM gauge singlet, all such interactions are higher dimensional operators.  Such
operators begin at dimension 7, though through electroweak symmetry breaking they also
imply effectively dimension 5 descendant operators as well.

The dimension 5 terms originate from,
\bea
\frac{1}{\Lambda^3_5} ~ \bar{\chi} \chi ~(D_\mu H)^{\dagger} D^\mu H
\eea
where $D_\mu H$ is the ordinary covariant derivative acting on the SM Higgs doublet.  Expanding out the covariant
derivative and replacing $H$ by its vacuum expectation value, we arrive at
\bea
\frac{m_W^2}{\Lambda_5^3} ~\bar{\chi} \chi ~W^{+ \mu} W^{-}_\mu
+ \frac{m_Z^2}{2 \Lambda_5^3} ~ \bar{\chi} \chi ~ Z^\mu Z_\mu ~.
 \eea
It is worth noting that while the overall size of both couplings may be varied by shifting
$v^2 / \Lambda^3_5$,  the ratio of the couplings to pairs of $W$ and $Z$ bosons
are fixed with respect to one other.  At higher order, this operator also results in
couplings to pairs of photons and to $Z \gamma$ through loops of $W$ bosons.

At dimension 7, there are also couplings to the kinetic terms of the electroweak bosons,
\bea
L= \frac{1}{\Lambda_7^3} ~\bar{\chi} \chi ~ \sum_i k_i  F_i^{\mu \nu} F^i_{\mu \nu}
\label{eq:SMSinglet}
\eea
\noindent
where $F_i$, $i=1,2,3$ are the field strengths for the SM $U(1)$, $SU(2)$, and $SU(3)$ gauge groups.
The couplings of dark matter to pairs of SM gauge bosons are given by:
\bea
g_{gg}&=&\frac{k_3}{\Lambda_7^3} \\
g_{WW}&=&\frac{2k_2}{s_w^2 \Lambda_7^3} \\
g_{ZZ} &=& \frac{1}{4 s_w^2 \Lambda_7^3} \left(\frac{k_1 s_w^2}{c_w^2}+\frac{k_2 c_w^2}{s_w^2} \right) \\
g_{\gamma\gamma}&=&\frac{1}{4 c_w^2}\frac{k_1+k_2}{\Lambda_7^3} \\
g_{Z\gamma} &=& \frac{1}{2 s_w c_w \Lambda_7^3} \left(\frac{k_2}{s_w^2}-\frac{k_1}{c_w^2} \right)
\label{eq:prefactors}
\eea
where $s_w$ and $c_w$ are the sine and cosine of the weak mixing angle, respectively.
For these kinetic operators, the over-all size can be though of as controlled by $k_2 / \Lambda_7^3$,
but there is still freedom to adjust the relative importance of various pairs by adjusting $k_1 / k_2$.

\section{Dark Matter Production in Association with a $Z$ Boson}

\begin{figure}
\includegraphics[width=1.5in]{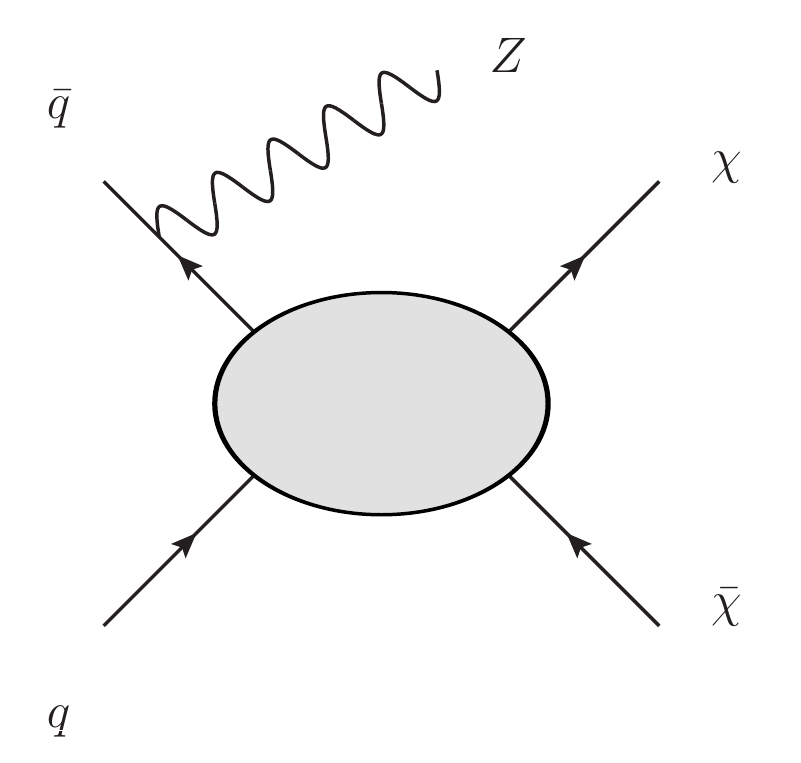}\\
\includegraphics[width=1.65in]{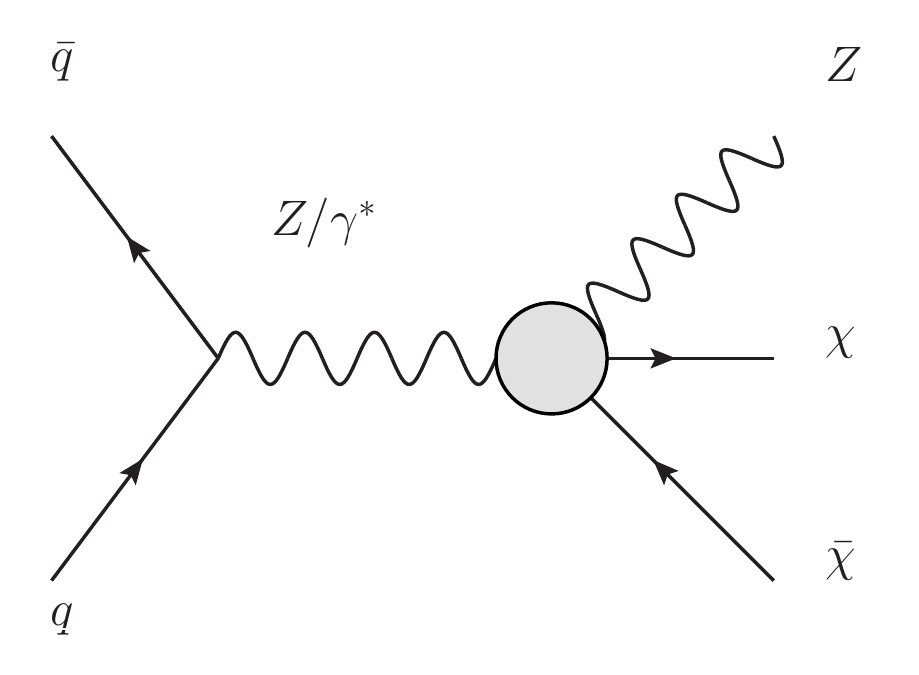}
\caption{Representative diagrams for production of dark matter pairs ($\chi \bar{\chi}$) associated with
a $Z$ boson in theories where dark matter interacts with quarks (top) or directly with $Z$ boson pairs (bottom).}
\label{diag}
\end{figure}

\begin{table}
\caption{Production cross sections (in fb) for pair production of WIMPs in
  association with a $Z$ boson,
  $pp\rightarrow Z\chi\bar{\chi} \rightarrow \ell^+ \ell^- \chi\bar{\chi}$, in theories where
  the dark matter interacts primarily with quarks, for $\Lambda_i = 1$~TeV and $\sqrt{s} = 7$~TeV. }
\label{tab:cs}
\begin{tabular}{lrrrr}
\hline\hline
$m_{\chi}$ (GeV) &  ~~~~~D1 & ~~~~~~~D5 & ~~~~~~~D8 & ~~~~~~~D9 \\
~ &  $[\times 10^{-8}]$ \\ \hline
$\le 10$ &	0.94 &	0.56 &	0.55 &	7.9 \\
~~~100 &	0.59 &	0.51 &	0.42 &	6.9 \\
~~~200 &	0.28 &	0.40 &	0.27 &	5.2 \\
~~~400 &	0.05 &	0.20 &	0.09 &	2.4 \\
~~~1000 & $3 \times 10^{-4}$ &	0.01 &	0.002 &	0.1 \\
\hline\hline
\end{tabular}
\end{table}

The process of interest is pair-production of dark matter particles in conjunction with one $Z$ boson.  Representative
Feynman diagrams are shown in Figure~\ref{diag}, for both the case of interactions with quarks as well as dark matter
which interacts directly with weak bosons.  In order to match on to the existing ATLAS $ZZ$ measurement, we
consider $pp$ collisions at $\sqrt{s}=7$ TeV.  Cross sections for each of the quark operators of Eq.~(\ref{eq:EFTq})
are presented for various dark matter masses and the corresponding $\Lambda$ set equal to 1~TeV
in Table~\ref{tab:cs}.
These cross sections scale as $\propto 1 / \Lambda^6_{\rm D1}$ for operator D1, and as $\propto 1 / \Lambda^4$ for
D5, D8, and D9.  The rates for D1 are considerably smaller, due to suppression of the contribution of valence
quark contributions by the small up and down quark masses.  For this operator, loops involving the top quark
are expected to be sizable and will result in an increase of the rate \cite{Haisch:2012kf}.  

In part because of the rather stiff branching ratio penalty in asking for $Z \rightarrow e^+ e^-$ or $\mu^+ \mu^-$,
the rate for the operators describing interactions with quarks is somewhat smaller than the corresponding rates
for mono-jets, mono-photons, or mono-$W$s.  However, the mono-$Z$ signature is nonetheless worth exploring,
in part because it samples a different weighting of couplings to up-type versus down-type quarks, but also because
the systematic uncertainties on the backgrounds should scale more favorably for mono-$Z$ than for mono-jets
or even mono-photons, given that fake ``QCD" backgrounds should be much smaller for mono-$Z$s.

For the case of direct interactions with weak bosons, the dimension 5 operator is mediated only by $Z$ exchange,
whereas the dimension 7 operator contains a mixture of $Z$ and photon exchange, with the relative importance of the
two controlled by $k_1 / k_2$.  We consider two example admixtures of the dimension 7 operators:
\begin{itemize}
\item $k_1 = k_2$ leading to contributions from both $Z$ and $\gamma$ exchange.
\item $k_1 = c_w^2 / s_w^2 k_2$, for which the $\gamma$ exchange graph is negligible.
\end{itemize}

\section{ATLAS results}

The ATLAS collaboration performed a measurement of the $ZZ\rightarrow
\ell\ell\nu\nu$ cross-section~\cite{atlaszz} in $pp$ data with
$\sqrt{s}=7$ TeV and integrated
luminosity of 4.6 fb$^{-1}$.
The fiducial region is defined as:
\begin{itemize}
\item two same-flavor opposite-sign electrons or muons, each with $p_{\rm
    T}^{\ell} > 20$ GeV, $|\eta^{\ell}|<2.5$;
\item dilepton invariant mass close to the $Z$ boson mass:
  $m_{\ell\ell} \in [76, 106]$ GeV;
\item no particle-level jet with $p^j_{\rm T} >$ 25 GeV and
  $|\eta^j|<$4.5;
\item $(|p_{\rm T}^{\nu\bar{\nu}} - p_{\rm T}^Z|)/ p_{\rm T}^Z < 0.4$;
\item $-p_{\rm T}^{\nu\bar{\nu}} \times \cos( \Delta\phi(p_{\rm
    T}^{\nu\bar{\nu}},p_{\rm T}^Z) ) > 75$ GeV.
\end{itemize}
The results are consistent with
Standard Model expectations (within $1\sigma$ for the $\mu^+ \mu^-$ channel and within
$2\sigma$ over-all), as shown in Table~\ref{tab:zzdata}.

\begin{table}
\caption{Expected backgrounds and observed data in the ATLAS
  $ZZ\rightarrow \ell\ell\nu\nu$ analysis~\cite{atlaszz}  in $pp$ collisions
  at $\sqrt{s}=7$ TeV with integrated luminosity of 4.6 fb$^{-1}$. The first uncertainty is statistical and systematic and 
  the second uncertainty is luminosity.}
\label{tab:zzdata}
\begin{tabular}{lrrr}
\hline\hline
 & ~~~~~~~~~$ee\nu\nu$\ \   & ~~~~~~~~~$\mu\mu\nu\nu$\ \   & ~~~~~~~~~~$\ell\ell\nu\nu$\ \  \\
\hline
Background\ \ \  &\ \  $20.8 \pm 2.7$  &\ \  $26.1 \pm 3.3$ &\ \  $46.9
\pm 5.5$ \\
SM $ZZ\rightarrow \ell\ell\nu\nu$ & $17.8 \pm 1.8$ & $21.6 \pm 2.2$ & $39.3\pm 4.0$ \\ \hline
Total & $38.6 \pm 3.8$ & $47.7 \pm 4.6$ & $86.2 \pm 7.2$\\
\hline
Data & 35 & 52 & 87 \\
\hline\hline
\end{tabular}
\end{table}

We use the expected background yield with uncertainties to calculate an upper limit on the number of events due to a new
source which could be present in the collected data. Using the CLs
method~\cite{cls1,cls2}, we find $N < 18.0$ at 90\% confidence limit (CL).   We convert
this to a limit on the cross section ($\sigma$) for new physics based on
$N = \sigma \times \epsilon \times \mathcal{L}$, where $\epsilon$ is the
fraction of new physics events which satisfy the selection
requirements.

To aid the reinterpretation of the result, ATLAS has divided their
calculation of $\epsilon = A_{ZZ} \times C_{ZZ}$ into two pieces: the fiducial
acceptance ($A_{ZZ}$),
the fraction of events which fall into a specified parton-level fiducial region,
and the reconstruction efficiency ($C_{ZZ}$), the fraction of events in the
fiducial region which satisfy the final selection.  We expect the
reconstruction efficiency $C_{ZZ}$ to be largely model independent, as
the fiducial region is chosen such that $C_{ZZ}$
 is determined by the detector performance for specific final state
objects rather than the production mechanism. Therefore, a calculation of the fiducial acceptance for a new
model is all that is needed for an estimate of the total efficiency
$\epsilon$.  In terms of the fiducial acceptance $A_{ZZ}$, the limits are
\bea
\sigma < \frac{N}{ A_{ZZ} \times C_{ZZ} \times \mathcal{L}}~,
\eea
which results at the $90\%$ CL in,
\bea
\sigma(90\%~ {\rm CL}) <  \frac{18.0}{ A_{ZZ}
  \times 0.679 \times 4.6\ \mathrm{fb}^{-1}}~.
\eea

\section{Interpretation}

\begin{figure}
\includegraphics[width=3in]{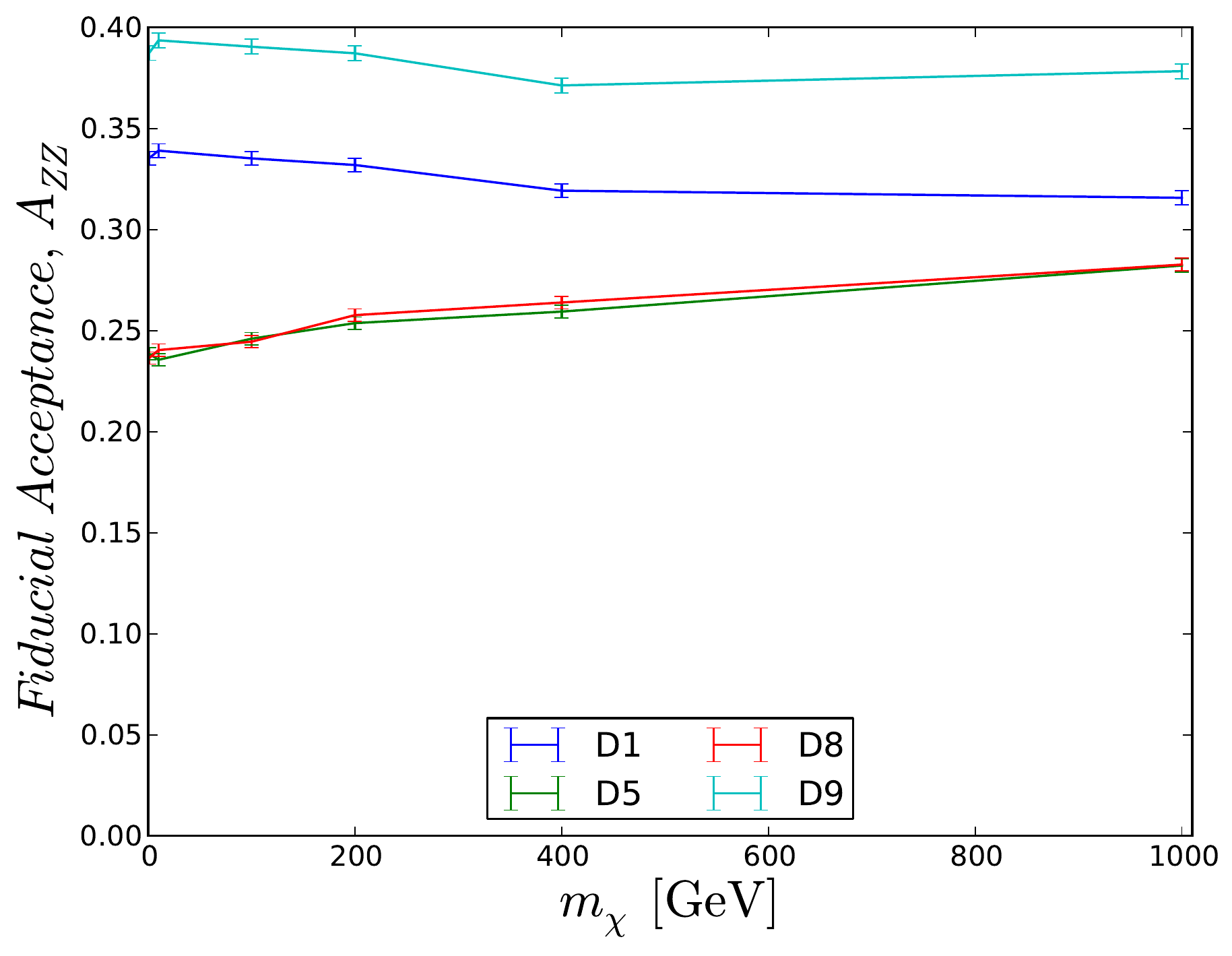}
\includegraphics[width=3in]{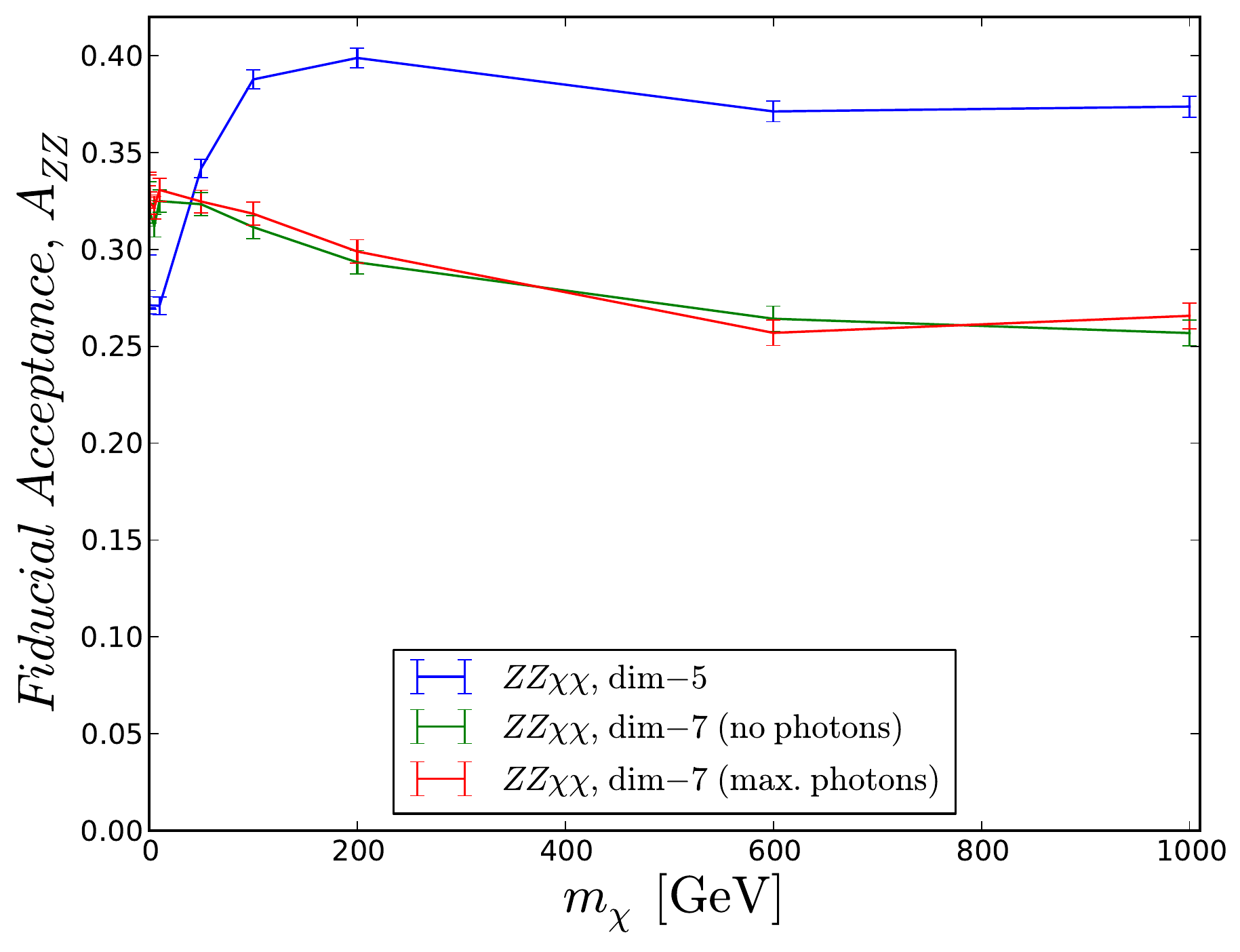}
\caption{Acceptance of the dark matter production process, $pp\rightarrow Z\chi\bar{\chi}$,
with $Z \rightarrow \ell^+ \ell^-$,
for the ATLAS $ZZ\rightarrow\ell\ell\nu\nu$ fiducial region (see text) as a function of the dark matter mass.}
\label{fig:acc}
\end{figure}

We simulate the predicted $Z \bar{\chi} \chi$ events for each of the EFTs described above, using
 {\sc madgraph}~\cite{madgraph}, with showering and
hadronization provided by {\sc pythia}~\cite{pythia} and particle-level jet
clustering with {\sc fastjet}~\cite{fastjet}.  
We work at tree level, though it should be noted that very recently the next to leading order
rates for several operators
have been computed \cite{Fox:2012ru}, and result in a modest increase in the expected rates.
For each of the EFTs and a variety of dark matter masses,
we compute the fiducial acceptance as defined above.  The resulting acceptances for each
EFT as a function of the dark matter mass are shown in Figure~~\ref{fig:acc}.  The acceptance varies
between about $\approx 20-35\%$, depending on the operator, and is roughly constant up to dark
matter masses of about 1 TeV.

\begin{figure}[t]
\includegraphics[width=3in]{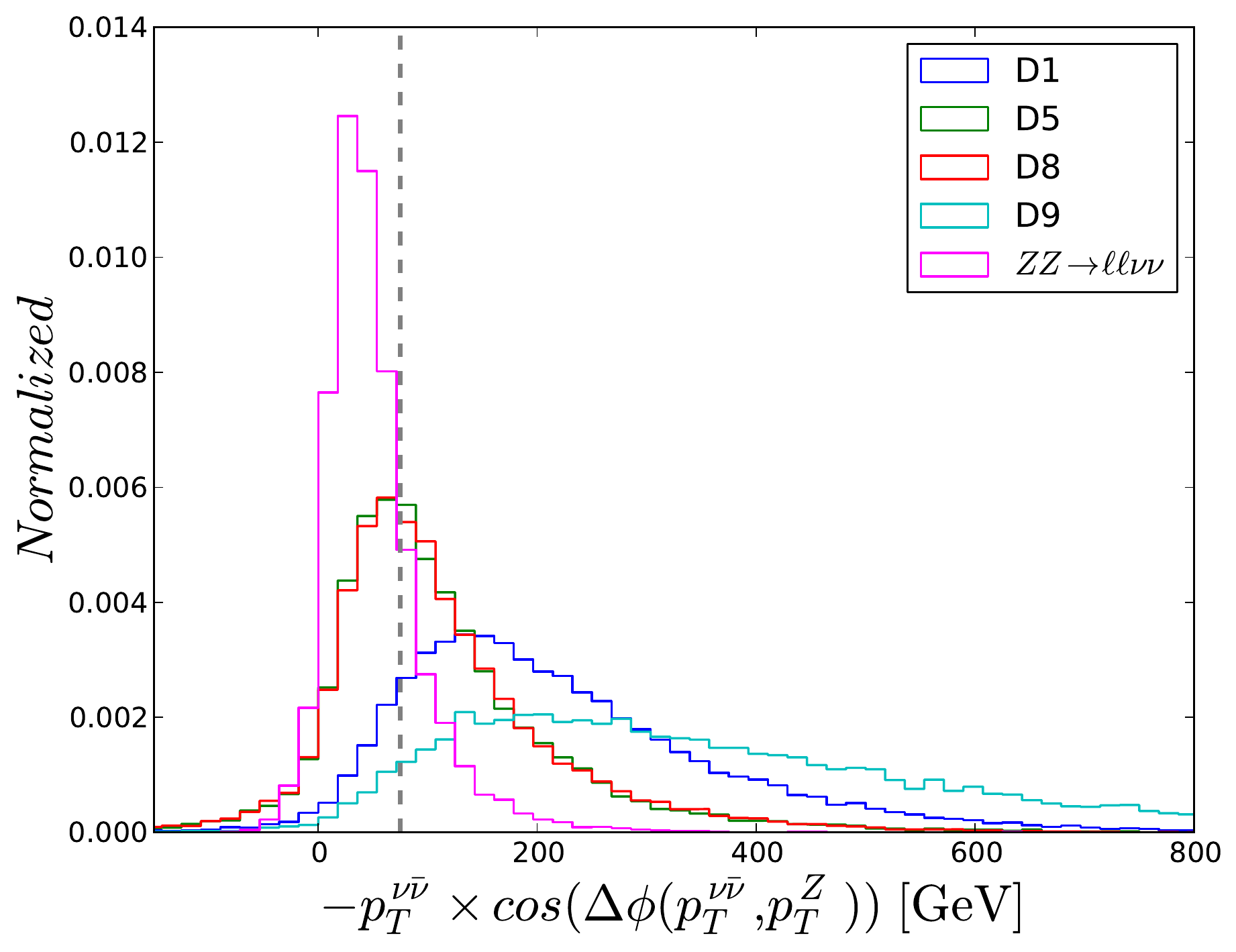}
\includegraphics[width=3in]{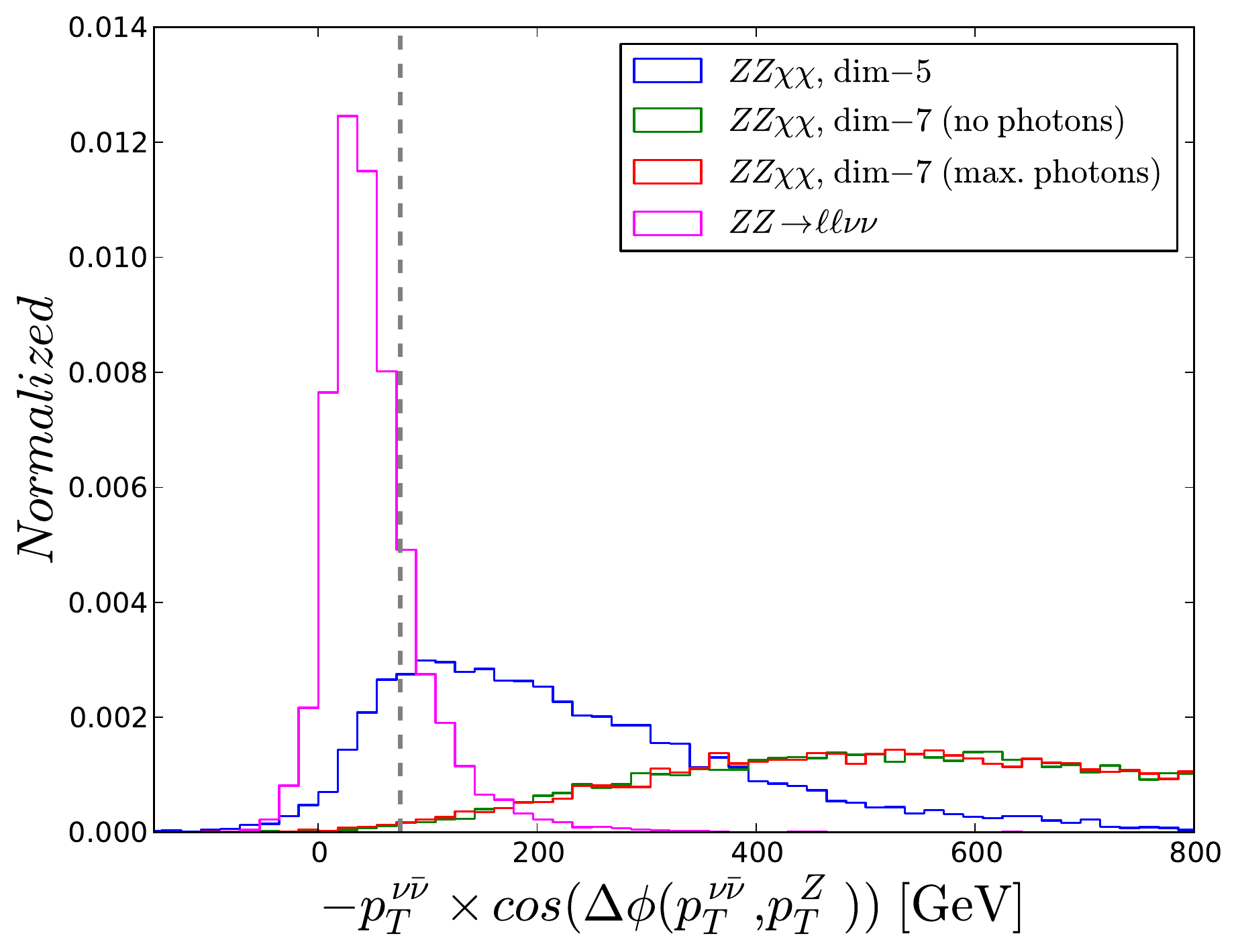}
\caption{  Distribution of axial missing transverse momentum in
  simulated Standard
  Model $ZZ\rightarrow\ell\ell\nu\nu$ and $pp\rightarrow Z\chi\bar{\chi}$, in the EFTs described in the text, for
  $\Lambda=1$ TeV and $m_\chi = 100$ GeV   at $\sqrt{s}=7$ TeV.}
\label{fig:met}
\end{figure}

\begin{figure}[t]
\includegraphics[width=3in]{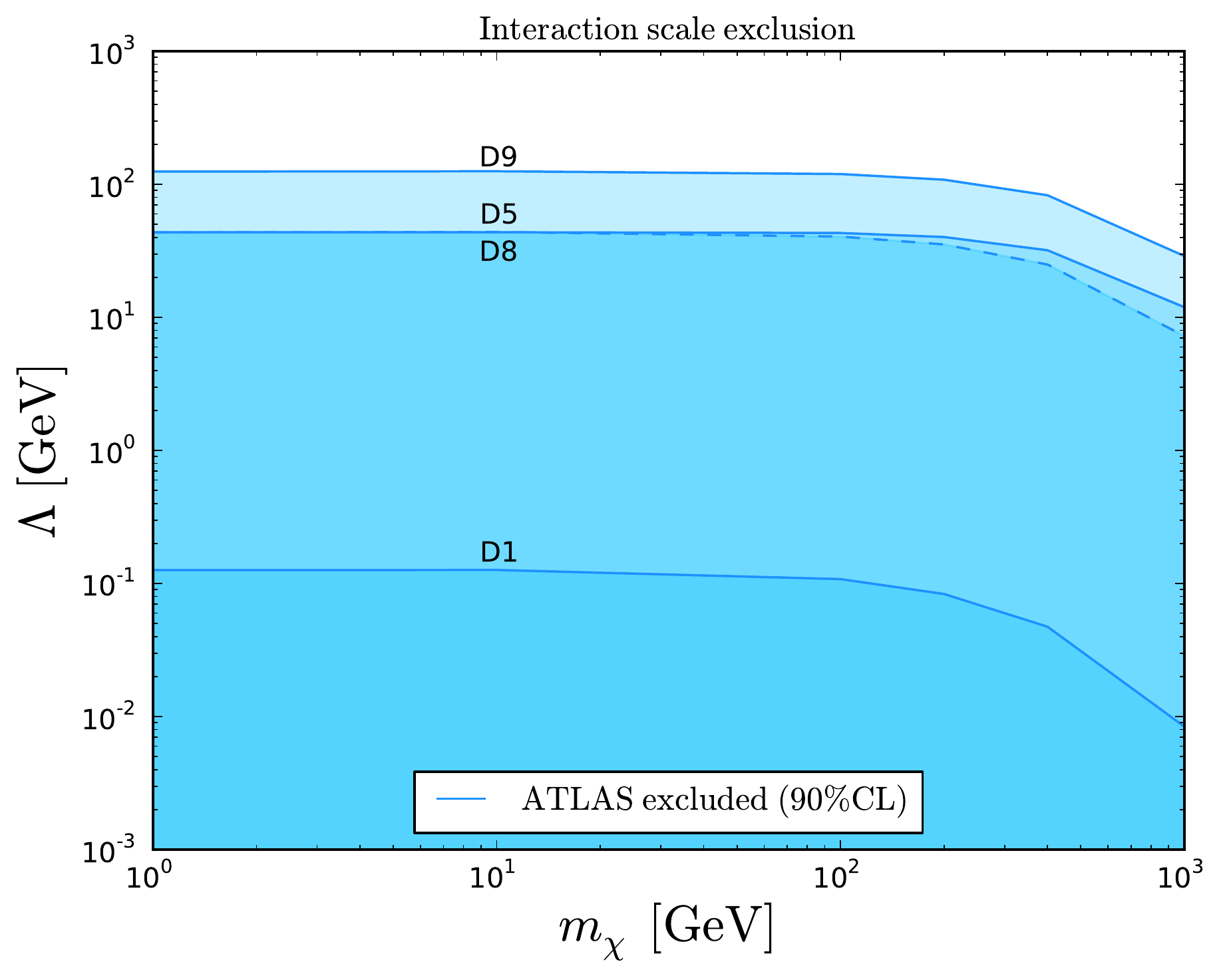}
\includegraphics[width=3in]{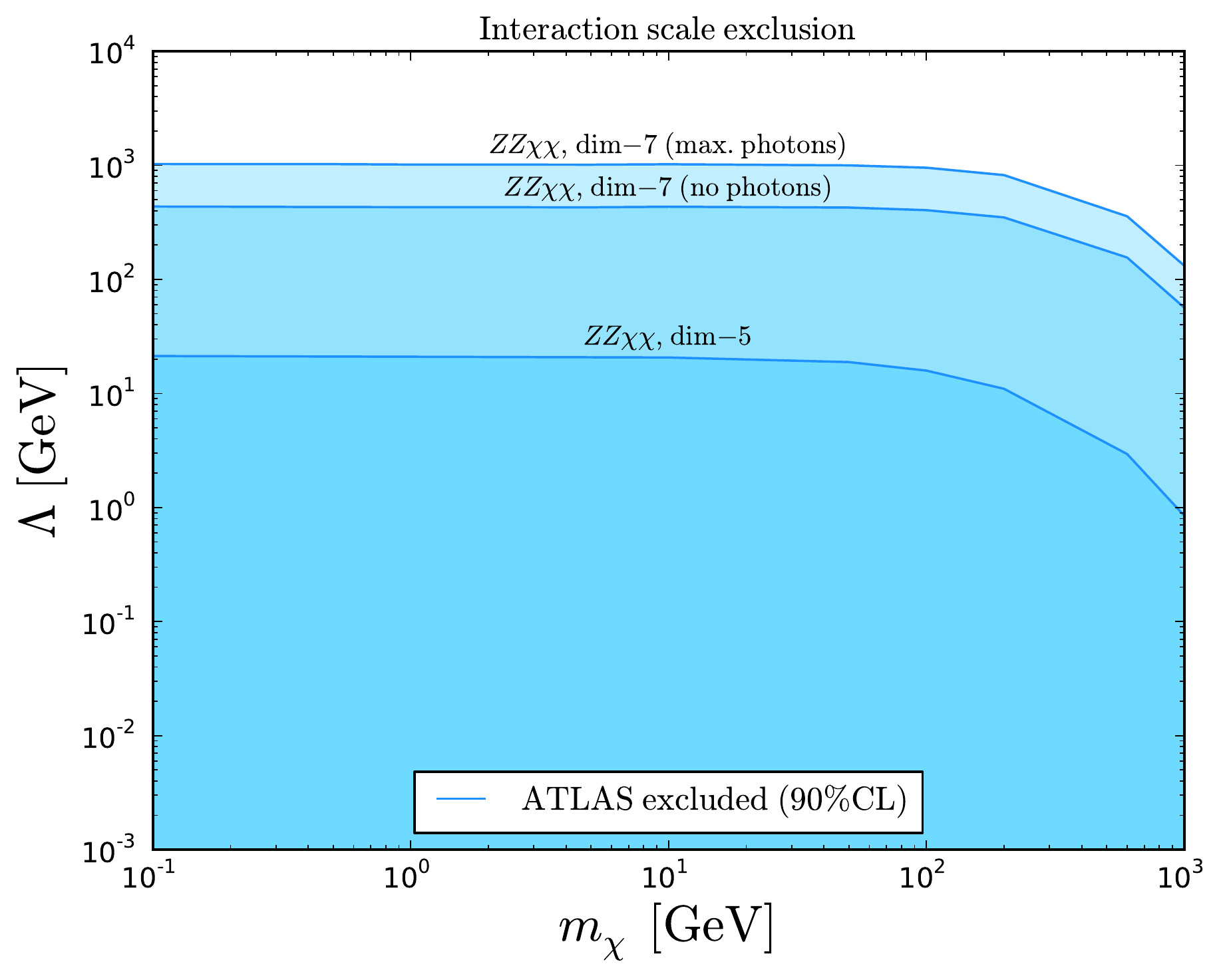}
\caption{ Excluded region (blue) at $90\%$ CL for the indicated interactions
as a function of dark matter mass.  The two regions for the dimension 7
$ZZ \bar{\chi} \chi$ correspond to the choices of $k_1 / k_2$ discussed in the
text, with either maximal contribution from photon graphs (upper curve)
or no contribution (lower curve).}
\label{fig:exclusions}
\end{figure}

Combined with the ATLAS measurement, the acceptances translate into bounds
on the production cross section of $\sigma(Z\chi\bar{\chi}) \lesssim 10-100$ fb at $90\%$~CL.
It is worth noting that the fiducial
acceptance is significantly higher for dark matter production than for the SM $ZZ$ production
due to the larger missing transverse momentum in the $Z\chi\bar{\chi}$, as shown in Figure~\ref{fig:met}
for a sample parameter point with $m_\chi = 100$~GeV and $\Lambda = 1$~TeV.  For each interaction
type, we determine the lower bound on the scale $\Lambda$ which characterizes its strength.  The results
are presented in Figure~\ref{fig:exclusions}, which indicate that for some types of interactions, scales on the
order of 100 GeV to TeV can be probed.  

\begin{figure}[t]
\includegraphics[width=3in]{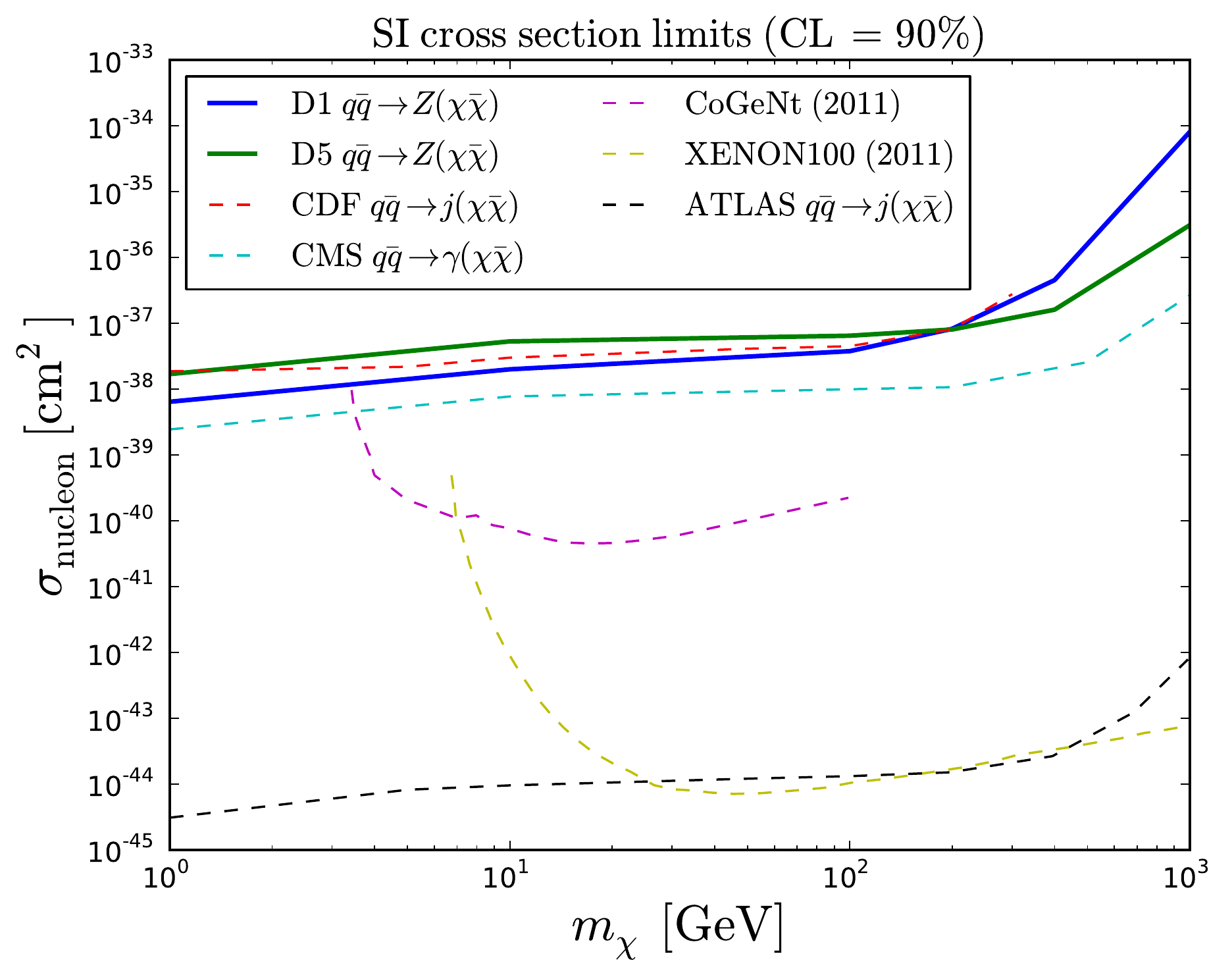} \includegraphics[width=3in]{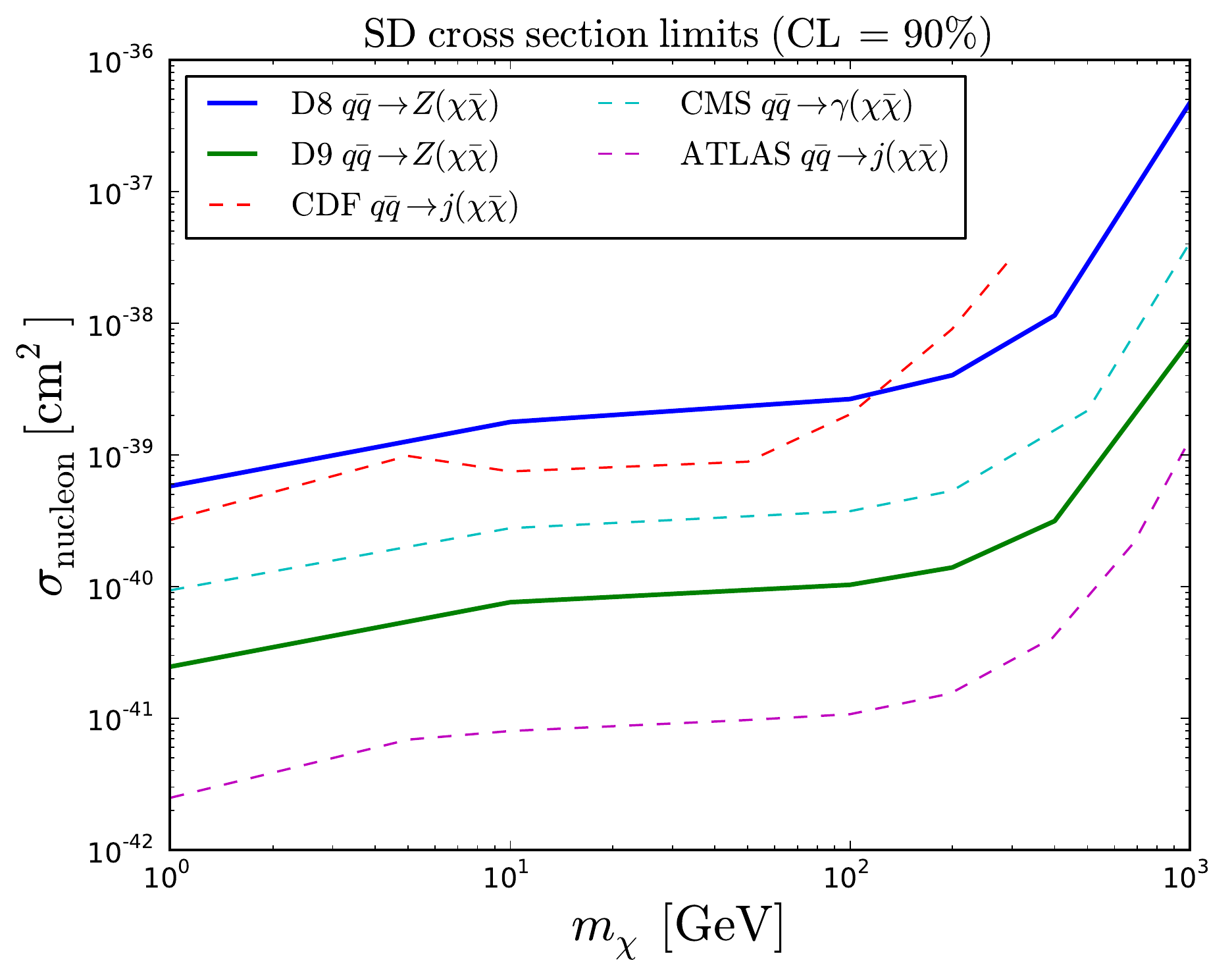}
\caption{ Exclusion regions at $90\%$ CL in the plane of the
dark matter mass ($m_\chi$) versus dark matter - nucleon spin-independent (top) or
  spin-dependent (bottom) cross-section plane.  Results from this analysis are
  compared to existing collider and direct detection limits.}
\label{fig:lim}
\end{figure}

For the case of interactions with quarks, we further
translate these bounds into the plane of spin-independent (SI) and spin-dependent (SD) scattering with
nucleons, as related in \cite{Goodman:2010ku}.  These bounds are shown in Figure~\ref{fig:lim}, along with bounds from CoGeNT \cite{Aalseth:2010vx} and
Xenon 100 \cite{Aprile:2012nq}, and competing collider results from mono-jet and mono-photon searches.  
As is typical, bounds from colliders provide a unique probe of very light dark matter particles,
and dominate as probes of spin-dependent interactions.  Of course, the collider bounds are subject to the
assumption that the EFT containing a contact interaction is a good description of the physics.  In cases with
light mediating particles, these bounds can sometimes be weaker.  Over-all, the two searches exhibit a
high degree of complementarity.

\section{Conclusions}

In this article, we have looked at the collider production of dark matter pairs in association with a $Z$ boson which decays
into charged leptons, a signature we refer to as a mono-$Z$.  
We work in the context of effective field theories in which the dark matter either interacts directly
with quarks, or with a pair of electroweak bosons.  We derive limits on the strength of such interactions based on the recent
ATLAS measurement of $ZZ$ production (where one $Z$ decays into charged leptons and the other into neutrinos) ands find
that the current limits already probe the TeV scale for some types of interactions.  

For the case of interactions directly with quarks, the mono-$Z$ signature provides limits which are somewhat weaker than those
from mono-jets or mono-photons.  Nonetheless, mono-$Z$ searches are expected to be less subject to systematic
uncertainties from jet energy scales and photon identification, and thus may scale better at large luminosities.  If a discovery
is made, the mono-$Z$ signature offers a different way to dissect the couplings of up-type versus down-type quarks.  If the dominant interaction is instead to pairs of
weak bosons, colliders offer a unique opportunity for discovery.

Our results illustrate the complementarity
between collider and direct searches of dark matter, and show how together they result in a more complete picture
of dark matter interactions with the SM fields.

\section{Acknowledgements}

DW, CS and AN are supported by grants from the Department of Energy
Office of Science and by the Alfred P. Sloan Foundation. TMPT
is supported in part by NSF Grant PHY-0970171.


\begin{thebibliography}{99}

\bibitem{Morrissey:2009tf}
  D.~E.~Morrissey, T.~Plehn and T.~M.~P.~Tait,
  Phys.\ Rept.\  {\bf 515}, 1 (2012)
  [arXiv:0912.3259 [hep-ph]].

\bibitem{cdfjmet}    CDF Collaboration,
  Phys.\ Rev.\ Lett.\  {\bf 108}, 211804 (2012).

\bibitem{ATLAS:2012ky}
  G.~Aad {\it et al.}  [ATLAS Collaboration],
  arXiv:1210.4491 [hep-ex].

\bibitem{Chatrchyan:2012me}
  S.~Chatrchyan {\it et al.}  [CMS Collaboration],
  JHEP {\bf 1209}, 094 (2012)
  [arXiv:1206.5663 [hep-ex]].

\bibitem{Aad:2012fw}
  G.~Aad {\it et al.}  [ATLAS Collaboration],
  arXiv:1209.4625 [hep-ex].

\bibitem{Chatrchyan:2012tea}
  S.~Chatrchyan {\it et al.}  [CMS Collaboration],
  Phys.\ Rev.\ Lett.\  {\bf 108}, 261803 (2012)
  [arXiv:1204.0821 [hep-ex]].

\bibitem{Bai:2012xg}
  Y.~Bai and T.~M.~P.~Tait,
  arXiv:1208.4361 [hep-ph].

\bibitem{atlaszz} 
 G.~Aad {\it et al.}  [ATLAS Collaboration],
  arXiv:1211.6096 [hep-ex] (2012).

\bibitem{ZprimeDM}
  F.~J.~Petriello, S.~Quackenbush, K.~M.~Zurek,
  arXiv:0803.4005v2 [hep-ph].

\bibitem{Bell:2012rg}
  N.~F.~Bell, J.~B.~Dent, A.~J.~Galea, T.~D.~Jacques, L.~M.~Krauss and T.~J.~Weiler,
  arXiv:1209.0231 [hep-ph].


\bibitem{Frandsen:2012db}
  M.~T.~Frandsen, U.~Haisch, F.~Kahlhoefer, P.~Mertsch and K.~Schmidt-Hoberg,
  JCAP {\bf 1210}, 033 (2012)
  [arXiv:1207.3971 [hep-ph]].




\bibitem{Beltran:2008xg}
  M.~Beltran, D.~Hooper, E.~W.~Kolb and Z.~C.~Krusberg,
  Phys.\ Rev.\ D {\bf 80}, 043509 (2009)
  [arXiv:0808.3384 [hep-ph]].

\bibitem{Shepherd:2009sa}
  W.~Shepherd, T.~M.~P.~Tait and G.~Zaharijas,
  Phys.\ Rev.\ D {\bf 79}, 055022 (2009)
  [arXiv:0901.2125 [hep-ph]].

\bibitem{Cao:2009uw}
  Q.~-H.~Cao, C.~-R.~Chen, C.~S.~Li and H.~Zhang,
  JHEP {\bf 1108}, 018 (2011)
  [arXiv:0912.4511 [hep-ph]].

\bibitem{Beltran:2010ww}
  M.~Beltran, D.~Hooper, E.~W.~Kolb, Z.~A.~C.~Krusberg and T.~M.~P.~Tait,
  JHEP {\bf 1009}, 037 (2010)
  [arXiv:1002.4137 [hep-ph]].

\bibitem{Goodman:2010yf}
  J.~Goodman, M.~Ibe, A.~Rajaraman, W.~Shepherd, T.~M.~P.~Tait and H.~-B.~Yu,
  Phys.\ Lett.\ B {\bf 695}, 185 (2011)
  [arXiv:1005.1286 [hep-ph]].

\bibitem{Bai:2010hh}
  Y.~Bai, P.~J.~Fox and R.~Harnik,
  JHEP {\bf 1012}, 048 (2010)
  [arXiv:1005.3797 [hep-ph]].

\bibitem{Goodman:2010ku}
  J.~Goodman, M.~Ibe, A.~Rajaraman, W.~Shepherd, T.~M.~P.~Tait and H.~-B.~Yu,
  Phys.\ Rev.\ D {\bf 82}, 116010 (2010)
  [arXiv:1008.1783 [hep-ph]].

\bibitem{Rajaraman:2011wf}
  A.~Rajaraman, W.~Shepherd, T.~M.~P.~Tait and A.~M.~Wijangco,
  Phys.\ Rev.\ D {\bf 84}, 095013 (2011)
  [arXiv:1108.1196 [hep-ph]].

\bibitem{Fox:2011pm}
  P.~J.~Fox, R.~Harnik, J.~Kopp and Y.~Tsai,
  Phys.\ Rev.\ D {\bf 85}, 056011 (2012)
  [arXiv:1109.4398 [hep-ph]].

\bibitem{Cheung:2012gi}
  K.~Cheung, P.~-Y.~Tseng, Y.~-L.~S.~Tsai and T.~-C.~Yuan,
  JCAP {\bf 1205}, 001 (2012)
  [arXiv:1201.3402 [hep-ph]].

\bibitem{Rajaraman:2012db}
  A.~Rajaraman, T.~M.~P.~Tait and D.~Whiteson,
  JCAP {\bf 1209}, 003 (2012)
  [arXiv:1205.4723 [hep-ph]].

\bibitem{Cotta:2012nj}
  R.~C.~Cotta, J.~L.~Hewett, M.~P.~Le and T.~G.~Rizzo,
  arXiv:1210.0525 [hep-ph].
  
\bibitem{Haisch:2012kf} 
  U.~Haisch, F.~Kahlhoefer and J.~Unwin,
  arXiv:1208.4605 [hep-ph].


\bibitem{cls1} {A. Read},   J. Phys. G: Nucl. Part. Phys. {\bf 28}, 2693 (2002);

\bibitem{cls2} {T. Junk},  Nucl. Instrum. Methods A {\bf 434}, 425
  (1999).


\bibitem{madgraph}
  J.~Alwall, M.~Herquet, F.~Maltoni, O.~Mattelaer and T.~Stelzer,
  JHEP {\bf 1106}, 128 (2011)
  [arXiv:1106.0522 [hep-ph]].

\bibitem{pythia}
  T.~Sjostrand, S.~Mrenna and P.~Z.~Skands,
  JHEP {\bf 0605}, 026 (2006)
  [hep-ph/0603175].

\bibitem{fastjet}
  M. Cacciari, G.P. Salam and G. Soyez, arXiv:1111.6097 (2011).

  
\bibitem{Fox:2012ru} 
  P.~J.~Fox and C.~Williams,
  arXiv:1211.6390 [hep-ph].
  

\bibitem{Aalseth:2010vx} 
  C.~E.~Aalseth {\it et al.}  [CoGeNT Collaboration],
  Phys.\ Rev.\ Lett.\  {\bf 106}, 131301 (2011)
  [arXiv:1002.4703 [astro-ph.CO]].

\bibitem{Aprile:2012nq} 
  E.~Aprile {\it et al.}  [XENON100 Collaboration],
  Phys.\ Rev.\ Lett.\  {\bf 109}, 181301 (2012)
  [arXiv:1207.5988 [astro-ph.CO]].
 
\end{thebibliography}
\end{document}